\documentclass{sig-alternate-05-2015}

\usepackage{tabularx}
\usepackage{wrapfig}
\usepackage{multirow}
\usepackage{hyperref}
\usepackage{float} 

\newcommand{\enquote}[1]{``#1''} 

\usepackage{tikz}
\newcommand\copyrighttext{%
  \footnotesize \textcopyright~ACM 2016. This is the author's version of the work. It is posted here for your personal use. Not for redistribution. The definitive Version of Record was published in Proceedings of the 9th International Workshop on Cooperative and Human Aspects of Software Engineering (CHASE '16). ACM, New York, NY, USA, 40-43, \url{https://doi.org/10.1145/2897586.2897602}.
}
\newcommand\copyrightnoticepreprint{%
    \begin{tikzpicture}[remember picture,overlay]
        \node[anchor=south,xshift=33.2em, yshift=10pt] at (current page.south) {\fbox{\parbox{\dimexpr\textwidth-\fboxsep-\fboxrule\relax}{\copyrighttext}}};
    \end{tikzpicture}%
}

\begin{document}

\CopyrightYear{2016}
\setcopyright{acmcopyright}
\conferenceinfo{CHASE'16,}{May 16 2016, Austin, TX, USA}
\isbn{978-1-4503-4155-4/16/05}\acmPrice{\$15.00}
\doi{http://dx.doi.org/10.1145/2897586.2897602}

\acmPrice{\$15.00}

\title{ScrumLint: Identifying Violations of Agile Practices Using Development Artifacts}

\numberofauthors{5} 
%
\author{
%
%
Christoph Matthies, Thomas Kowark, Keven Richly, \\ Matthias Uflacker, and Hasso Plattner\\
       \affaddr{Hasso Plattner Institute, University of Potsdam}\\
       \affaddr{August-Bebel-Str. 88}\\
       \affaddr{Potsdam, Germany}\\
       \email{\{firstname.lastname\}@hpi.de}
}
\date{16 May 2016}

\maketitle

\begin{abstract}
Linting tools automatically identify source code fragments that do not follow a set of predefined standards.
Such feedback tools are equally desirable for \enquote{linting} agile development processes.
However, providing concrete feedback on process conformance is a challenging task, due to the intentional lack of formal agile process models.
In this paper, we present ScrumLint, a tool that tackles this issue by analyzing development artifacts.
On the basis of experiences with an undergraduate agile software engineering course, we defined a collection of process metrics.
These contain the core ideas of agile methods and report deviations.
Using this approach, development teams receive immediate feedback on their executed development practices.
They can use this knowledge to improve their workflows, or can adapt the metrics to better reflect their project reality.
\end{abstract}

%
%

%
%

%
%

\vspace{-0.5cm}
\copyrightnoticepreprint



\section{Introduction}
The Unix utility \emph{lint}~\cite{Johnson1978}, a static code analysis tool, flags suspicious programming constructs in C source code.
It allows insights into problematic constructs in a fast and automated fashion.
\emph{ScrumLint} applies this approach to Scrum and agile processes.
It analyzes development artifacts, such as commits, testing statistics or user stories, and identifies patterns that constitute problems in the implementation of agile practices.
The tool was developed in the context of a university software engineering course, introducing undergraduate students to Scrum.
In line with teaching recommendations~\cite{JointTask2013}, the course features a hands-on software development project, which all students work on collaboratively.
This setup requires frequent feedback by the teaching staff to allow students to learn and adapt their processes quickly.
Yet, contrary to theoretical foundations, whose understanding can be assessed through exams, the quality of practical application is difficult to monitor~\cite{Igaki2014}.
One solution is to employ tutors, who are present during all Scrum meetings of teams~\cite{icse2016}.
They are able to gauge collaboration and can give immediate feedback.
This approach falls short, however, during the crucial teamwork phases, when actual programming takes place and urgent communication and organisational challenges arise.
Here it is still difficult to obtain information to base feedback on.
Instead of introducing a more controlled setting, which takes away from the core agile experience~\cite{Devedzic2011}, ScrumLint analyzes the development artifacts that are produced during regular development activities.
Already in medium size projects, such as our software engineering course with 40 participants, large amounts of development data is created.
The latest installment produced 379 user stories with 4707 revisions and 1802 commits featuring 26503 file changes.
As such, manually finding areas of improvement, where team members deviated from agile practices, is cumbersome and scales poorly with the amount of active participants.
ScrumLint automates this process, allowing insights into the state of implementation of agile practices in a team and provides a constantly available resource of feedback for team members.
It is publicly available under the MIT license\footnote{\url{https://github.com/chrisma/ScrumLint}}.

\section{ScrumLint Overview}

ScrumLint aims at supporting a development team in adopting or adhering to agile practices.
It identifies and quantifies \emph{violations}, instances where the executed process deviates from the defined one, as mandated by Scrum or agile best practices.
While the absence of detected violations does not imply a perfectly executed process, similar to linters or test coverage tools, identified violations can reveal problem areas.
These represent starting points for further analysis and discussion, activities that involve collaboration and communication between team members.

ScrumLint operates on aggregated development data collected from multiple sources, e.g. code repositories like Github or build logs from continuous integration services like Travis CI.
It applies a set of rules, referred to as \emph{conformance metrics}, to this data.
These metrics include information about the agile practices that are measured, as well as the specifics of how to measure and evaluate deviations.
We derived metrics from best practices and experiences based on running our software engineering course over the last five years, as well as literature~\cite{Sletholt2012}.
The main challenge lies in defining and formalizing the conformance metrics, the associated agile practices and the patterns that point to a violation.

\subsection{Conformance Metric Lifecycle}
\begin{figure}[htb]
	\centering
	\includegraphics[width=0.85\columnwidth]{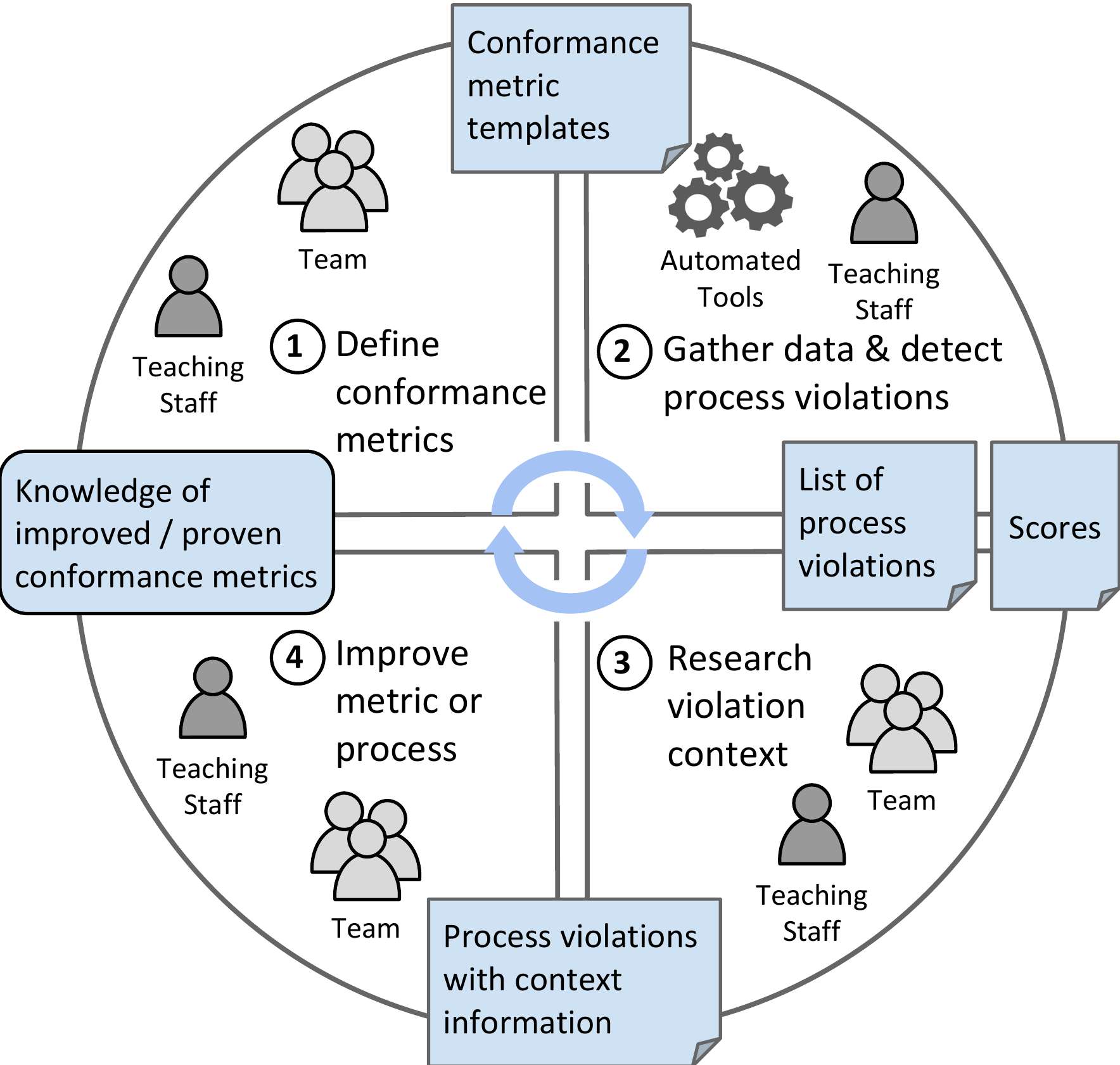}
	\caption{The conformance metric lifecycle.}
	\label{fig:lifecycle}
\end{figure}

We adapted Zazworka et al's.~\cite{zazworka2010developers} model of process nonconformance as a basis for detecting process violations (see Figure~\ref{fig:lifecycle}).

It is an iterative approach that demands that metrics continually undergo \enquote{improvement} steps to make sure they stay relevant to the project's changing context.
After a set of metrics and associated agile practices is defined, they are executed on the collected data, producing a list of violations.
The context of these violations needs to be assessed in order to determine the best cause of action.
For example, the changeset of a commit can be inspected to determine whether the connected violation is a false positive (which requires a change in the metric), or a true positive (which requires a change in process execution).
The desired changes to the system and the executed process are applied and the cycle begins anew.

\subsection{Result Presentation}
\begin{figure}[!htb]
	\centering
	\includegraphics[width=\columnwidth]{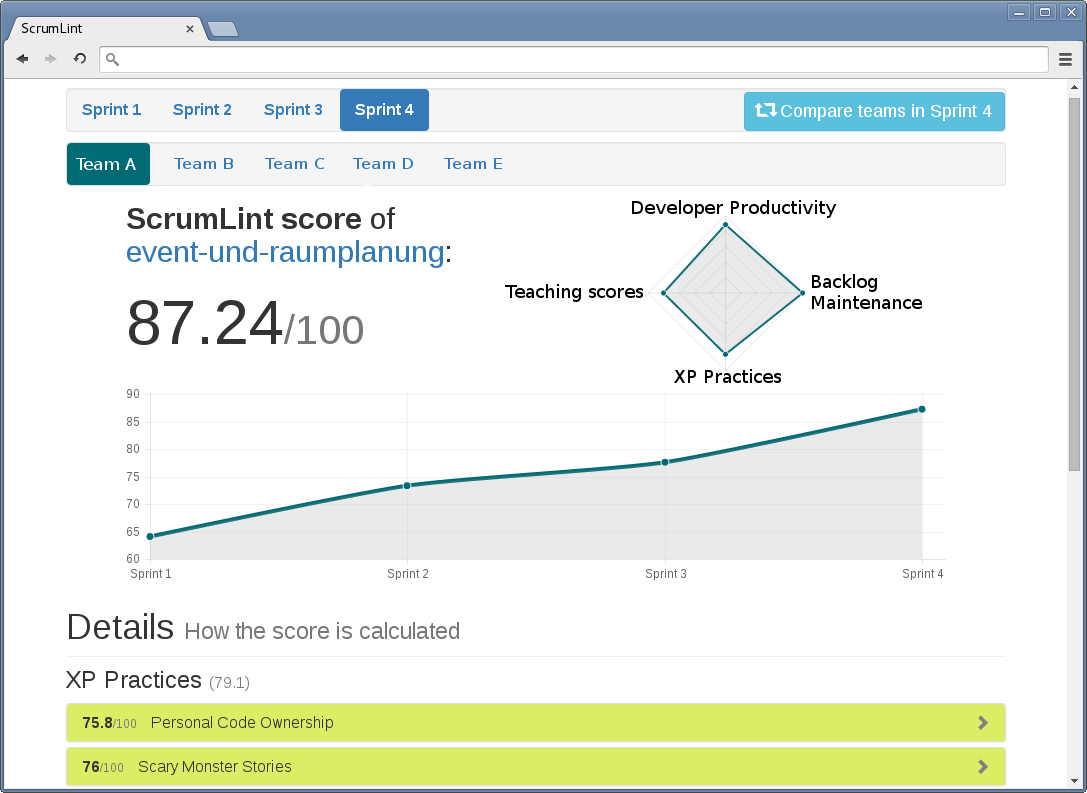}
	\caption{Screenshot showing the development of the ScrumLint score for a team.}
	\label{fig:teamoverview}
\vspace{-0.3cm}
\end{figure}

As ScrumLint is web-based, the output presented to users is a web page containing the identified violations and their details.
These are organized into categories, giving an overview of what process areas require attention.
Identified violations are visualised using line and radar charts, and a score, reflecting the severity of violations, is assigned to each metric.
These individual scores are aggregated into an overall \emph{ScrumLint score}, which represents the severity of all violations of agile processes in a team for a sprint.
It allows comparing teams' process conformance against each other and over iterations (see Figure~\ref{fig:screenshots}a).
A perfect score indicates that no violations were found while a low one indicates that the defined practices were rarely followed.
A screenshot of ScrumLint showing a team overview page is given in Figure~\ref{fig:teamoverview}.

\subsection{Related Approaches}
With the presented approach, ScrumLint is in line with recent similar tools, such as SQA-Mashup~\cite{Brandtner14} or Microsoft CodeFlow Analytics~\cite{Bird:2015aa}.
These also aim at providing easy-to-grasp overviews of potential problems in software engineering processes and allow their users to zoom in on concrete artefacts.
ScrumLint's main contribution, however, is that it is the first tool that aims to capture the core aspects of agile processes, in particular Scrum, and support teams beyond the standard agile metrics, such as Burndown charts and velocity calculation.



\section{Use Case}

In an agile development team, all team members should strive to adhere to agile practices.
A role in every team that is especially concerned with this is the Scrum Master (SM).
The SM is tasked with supporting her team, removing blockers, and suggesting improvements to the process.
She participates in development activities and has insights into how well her team is doing.
However, she does not have knowledge of how her team's implementation of Scrum compares to the other teams in the project.
This is of interest in order to find those areas of the process that other teams fared better and where there is learning potential.

\begin{figure*}[!htb]
	\centering
	\includegraphics[width=\textwidth]{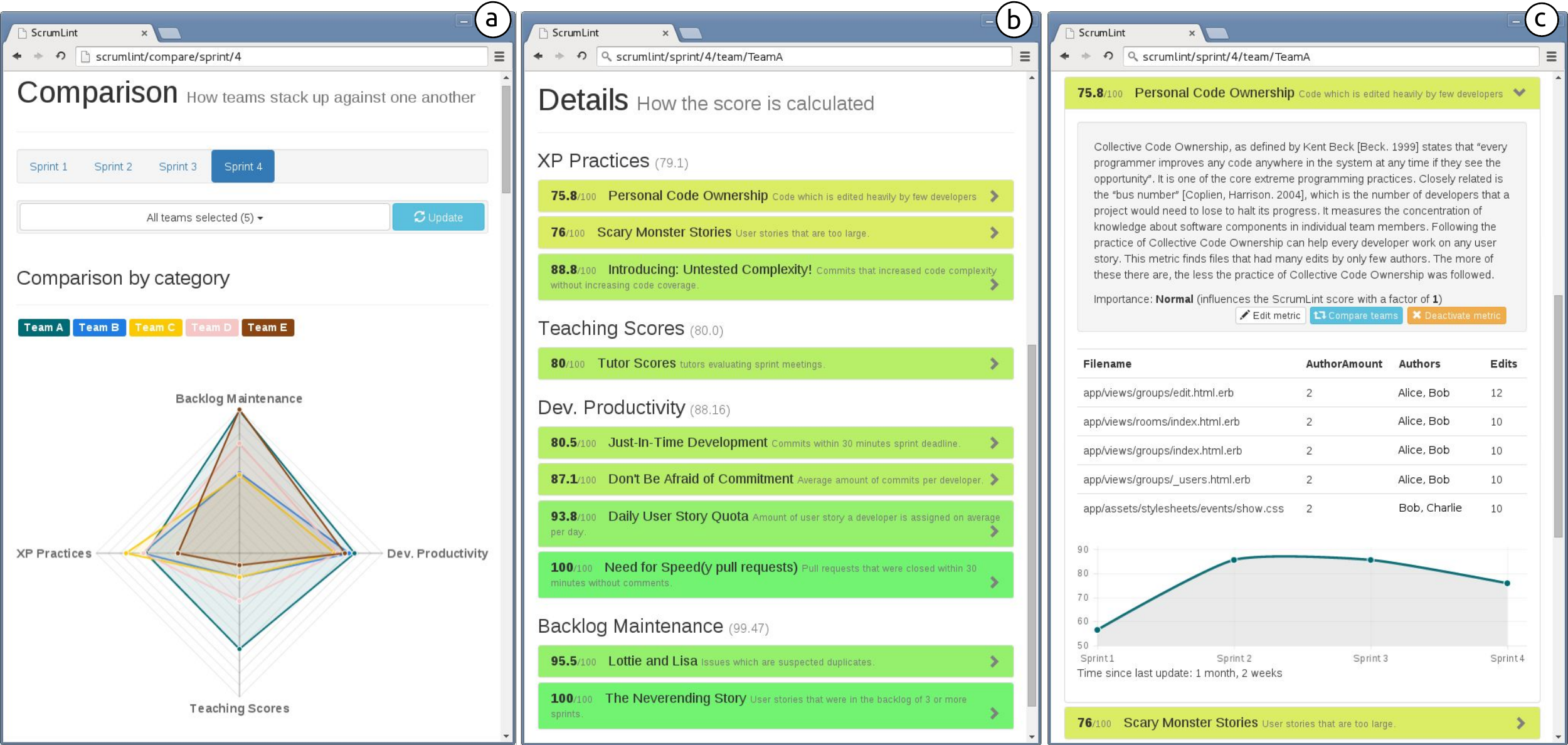}
	\caption{ScrumLint screenshots. Radar chart comparing teams by categories (a). List of conformance metrics ordered by their scores (b) and a specific metric's details expanded (c).}
	\label{fig:screenshots}
\vspace{-0.3cm}
\end{figure*}

This is a prime example where ScrumLint can be employed.
It can support the SM in the following tasks:

\vspace{0.2cm}

\textbf{Identify category}.
The SM starts research by visiting the team comparison view.
Its radar chart compares all teams by category scores (see Figure~\ref{fig:screenshots}a).
The SM is able to identify categories where her team scored significantly higher or lower compared to other teams.
Scoring lower might indicate a problem, while scoring higher could mean the team tried something new that is useful to other teams as well.
The SM notices that her Team scored lower than other teams in the \emph{Backlog Maintenance} category.

\vspace{0.05cm}

\textbf{Identify metric}.
Next, the SM heads to the detail section of the team-centric view for the last sprint (see Figure~\ref{fig:teamoverview}).
Here, all categories and the metrics within, are listed, sorted by metric scores (see Figure~\ref{fig:screenshots}b,c).
She selects the metric at the top of the Backlog Maintenance section, \emph{The Neverending Story}, which received the lowest score in this category.

\vspace{0.05cm}

\textbf{Identify artifacts, research context}.
The details of user stories that were in the last sprint backlog as well as in the two previous ones, are presented.
By following the links, the SM is led to the concrete story on Github and reads its details.

\vspace{0.05cm}

\textbf{Enact improvements}.
Judging from the posted comments and the size of the user story, she concludes that the story is too large to be completed by the team in one iteration.
She attends the next Scrum meeting, pointing out the identified stories to the assembled team and consulting with them on improving the executed process on the basis of the concrete data.
Furthermore, teams that did well on this metric can be involved to find out what has worked well for them, e.g. splitting user stories by the create, read, update and delete aspects.
With the knowledge of what concrete issues should be tackled and the ability to track metrics during the sprint, the team can improve their process in the next iteration.

\section{Architecture}
\begin{figure}[H]
	\centering
	\includegraphics[width=0.9\columnwidth]{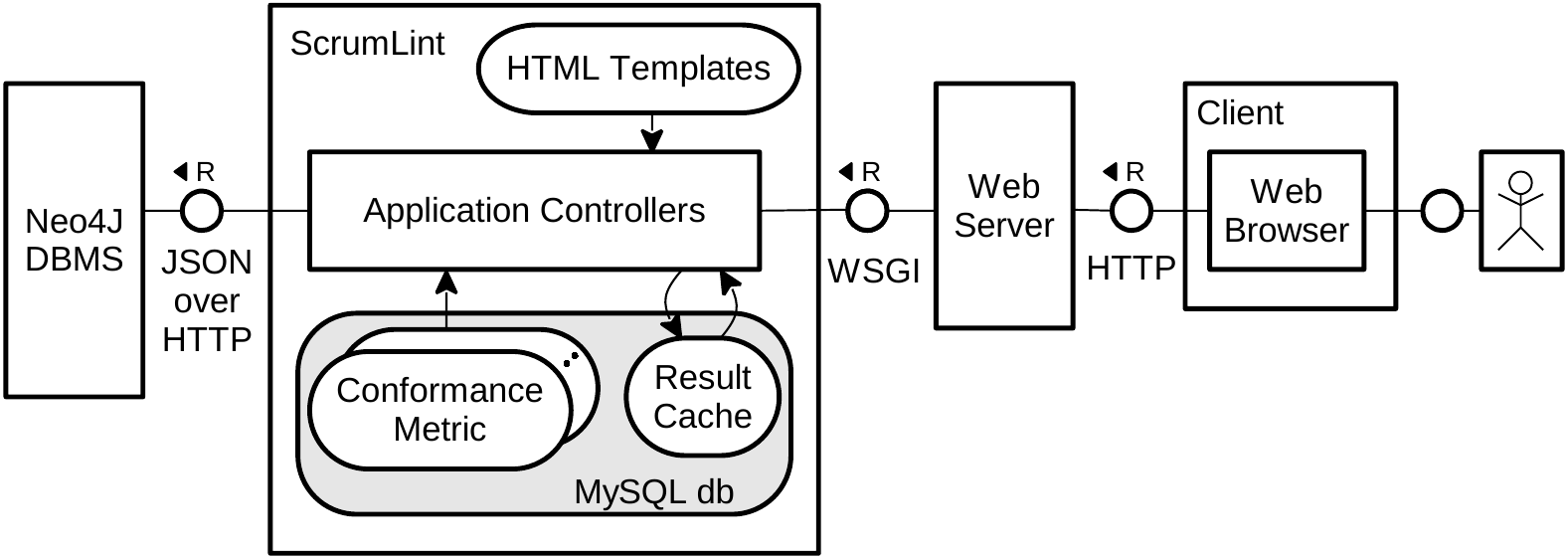}
	\caption{FMC block diagram of the architecture of ScrumLint.}
	\label{fig:slarch}
\vspace{-0.2cm}
\end{figure}

ScrumLint is written in Python using the Django framework (see Figure \ref{fig:slarch}).
It implements models for conformance metrics and calculates them based on development artifacts, which are stored in a Neo4j graph database.
Results of metrics are cached within the application and are updated in configurable intervals.

\subsection{Data Collection}
Collection and storage of development artifacts are separated from ScrumLint, in order to simplify its reuse in different collaboration infrastructures.
Currently, development artifacts from Github (commits, milestones, issues) as well as test run statistics and complexity measures for each commit are collected and written to the graph database.
Furthermore, information on sprints and the composition development teams are extracted from Github.
We employ a custom solution for this task, but standard solutions such as \emph{SonarQube}~\cite{campbell2013sonarqube} could easily be adapted.
Adding additional data sources involves creating a new importer that has knowledge of how to the source data is connected with the existing data and that is able to write it to the database.
Conformance metrics can then take advantage of the newly available data.

\subsection{Conformance Metrics}
Currently, the system includes ten different conformance metrics, in the categories \enquote{XP Practices}, \enquote{Backlog Maintenance} and \enquote{Developer Productivity}.
These measure details of the Scrum process that students had problems adopting in the last iterations of the course.
Table~\ref{table:backlogmaintenance} gives an overview of the metrics of the Backlog Maintenance category.
\begin{table}
\centering
\caption{Conformance metrics of the Backlog Maintenance category.}
\label{table:backlogmaintenance}
\begin{tabularx}{\columnwidth}{|p{2.5cm}|X|} \hline
\textbf{Name} & \textbf{Summary}\\ \hline
The Neverending Story & User Stories in multiple backlogs.\\ \hline
Monster Stories & Unusually large User Stories.\\ \hline
Lottie and Lisa & Suspected duplicate User Stories.\\ \hline
\end{tabularx}
\vspace{-0.3cm}
\end{table}
In order to execute a conformance metric, it must contain two main parts: a query that extracts the violation instances, and a rating function which calculates the corresponding score.
Queries are defined using Cypher\footnote{{\scriptsize \url{http://neo4j.com/docs/stable/cypher-query-lang.html}}}, the query language used by Neo4j, and need to include placeholders for identification of sprints and, if necessary, teams.
Thus, the system is able to run each query for all teams and sprints separately to calculate score changes over time.
Table \ref{table:confmetric} shows an example of the most important features of such a metric.


\begin{table}
\centering
\caption{Excerpt of \enquote{The Neverending Story} conformance metric.}
\label{table:confmetric}
\begin{tabularx}{\columnwidth}{|X|} \hline
{\bf Name:} The Neverending Story\\ \hline
{\bf Category:} Backlog Maintenance\\ \hline
{\bf Severity:} High\\ \hline
{\bf Data source:} User story tracker\\ \hline
{\bf Description:} Ideally, a sprint backlog contains exactly as many user stories as the team can complete in the iteration [Schwaber, 2013] \ldots\\ \hline
{\bf Query:} \newline
\texttt{MATCH} (e:Event)-[:issue]-(i:Issue)-[:labels]-(l:Label) \newline
\texttt{WHERE} e.event=\enquote{milestoned} \texttt{AND} e.title \texttt{IN} [\textit{\{sprint\_list\}}] \texttt{AND} l.name = \enquote{\textit{\{team\}}} \newline \texttt{WITH} i, collect(\texttt{DISTINCT} e.milestone\_title) as Sprints \newline \texttt{WITH} i, Sprints, length(Sprints) as InSprints \newline \texttt{WHERE} InSprints > \textit{\{threshold\}} \newline \texttt{RETURN} i as Issues, InSprints, Sprints\\ \hline
{\bf Rating function:} $max(0, 100-(\frac{\#violations}{\#totalUS}*100*AvgInSprints))$, where $\#violations$ = amount of query results, $\#totalUS$ = length of Sprint Backlog, $AvgInSprints$ = average amount of sprint backlogs the violations were in.\\ \hline
\end{tabularx}
\vspace{-0.3cm}
\end{table}


In order to add another metric, a new instance of a conformance metric is created and the necessary fields are filled.

Users can adapt queries to their own project setup through an administrative user interface.
First, the severity of a metric, the factor that a single metric influences the overall score with, can be changed.
Second, what pattern is extracted as a violation can be adapted by changing the database query directly.
Third, the rating function that calculates a score from violations, can be adapted.
For example, thresholds in the formula can be changed, or a new exponential model can replace a linear one, where a small increase in violations result in a drastically reduced score.


\section{Conclusion}

ScrumLint allows executing and visualizing a collection of conformance metrics for a given project.
It explicitly takes into account agile concepts such as user stories, working in agile development teams, and iterations.
Results are grouped by iterations, which allows comparing conformance to Scrum practices over time.
ScrumLint fits into the existing Scrum cycle, e.g. by supporting Sprint Retrospectives at the end of sprints.
As most of the implemented metrics rely on existing development artifacts, existing workflows do not need to change.
ScrumLint can alleviate the need to manually analyze development data, allowing the focus on the identified problem areas of the process.
Violations can be tracked down to the actual artifact, e.g. a user story, allowing discussions on the basis of concrete data.

As conformance metrics are the basis of ScrumLint, their quality is mainly responsible for the quality of overall results.
However, there is little research yet on what constitutes best practices for agile metrics.
Zazworka et al. state that the \enquote{biggest challenge was to find definitions for the XP practices that contained enough detail}~\cite{zazworka2010developers}.
We were able to define metrics for common Scrum implementation issues based on experiences gathered from running our undergraduate software engineering course over the last years.
Using ScrumLint and this relatively small amount of metrics, we were able to extract areas of improvement in the Scrum workflow of student teams for all iterations of the project.
We're now interested in employing our tool in a professional setting with refined and extended metrics.

%
\bibliographystyle{abbrv}
\bibliography{library} 


\end{document}